\documentclass[10pt,twocolumn,letterpaper]{article}

\usepackage{iccv}
\usepackage{times}
\usepackage{epsfig}
\usepackage{graphicx}
\usepackage{amsmath}
\usepackage{amssymb}
\usepackage{comment}
\usepackage{enumitem}
\usepackage{caption}
\usepackage{subcaption}
\usepackage{multirow}
\usepackage{tabularx}
\usepackage{booktabs}
\newcolumntype{Y}{>{\centering\arraybackslash}X}

\usepackage[breaklinks=true,bookmarks=false]{hyperref}

\iccvfinalcopy 

\def\iccvPaperID{****} 

\ificcvfinal\fi

\begin{document}

\title{A Simple and Robust Framework for Cross-Modality Medical Image Segmentation applied to Vision Transformers}

\author{Matteo Bastico\textsuperscript{1}, David Ryckelynck\textsuperscript{1}, Laurent Cort\'e\textsuperscript{1}, Yannick Tillier\textsuperscript{2}, Etienne Decenci\`ere\textsuperscript{3}\\
Mines Paris, Université PSL\\
\textsuperscript{1}Centre des Matériaux (MAT), UMR7633 CNRS, 91003 Evry, France\\
\textsuperscript{2}Centre de Mise en Forme des Matériaux (CEMEF), UMR7635 CNRS, 06904 Sophia Antipolis, France\\
\textsuperscript{3}Centre de Morphologie Mathématique (CMM), 77300 Fontainebleau, France\\
{\tt\small \{matteo.bastico, etienne.decenciere\}@minesparis.psl.eu} \\
}
\maketitle
\ificcvfinal\thispagestyle{empty}\fi

\begin{abstract}
   When it comes to clinical images, automatic segmentation has a wide variety of applications and a considerable diversity of input domains, such as different types of Magnetic Resonance Images (MRIs) and Computerized Tomography (CT) scans. This heterogeneity is a challenge for cross-modality algorithms that should equally perform independently of the input image type fed to them. Often, segmentation models are trained using a single modality, preventing generalization to other types of input data without resorting to transfer learning techniques. Furthermore, the multi-modal or cross-modality architectures proposed in the literature frequently require registered images, which are not easy to collect in clinical environments, or need additional processing steps, such as synthetic image generation. In this work, we propose a simple framework to achieve fair image segmentation of multiple modalities using a single conditional model that adapts its normalization layers based on the input type, trained with non-registered interleaved mixed data. We show that our framework outperforms other cross-modality segmentation methods, when applied to the same 3D UNet baseline model, on the Multi-Modality Whole Heart Segmentation Challenge. Furthermore, we define the Conditional Vision Transformer (C-ViT) encoder, based on the proposed cross-modality framework, and we show that it brings significant improvements to the resulting segmentation, up to 6.87\% of Dice accuracy, with respect to its baseline reference. The code to reproduce our experiments and the trained model weights are available at \url{https://github.com/matteo-bastico/MI-Seg} and in the Supplementary Code.
\end{abstract}

\section{Introduction}
\begin{figure}[t]
    \centering
    \includegraphics[width=0.47\textwidth]{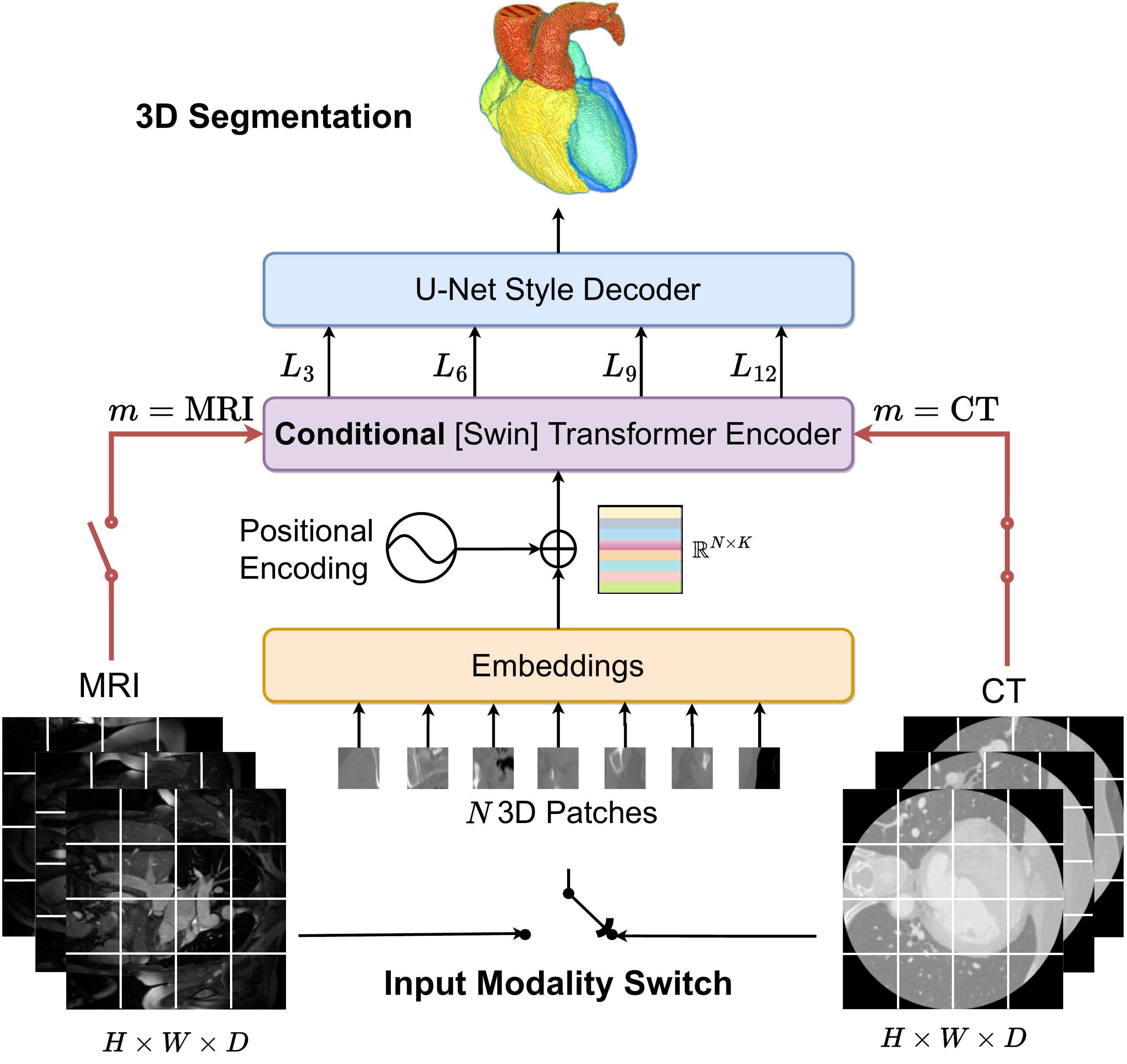}
    \caption{Overview of the proposed cross-modality clinical image segmentation technique applied to the Swin-UNETR segmentation model \cite{hatamizadeh_swin_2022}. The input modality can be arbitrarily switched to obtain the desired segmentation, while keeping the model unchanged. In conditional models, the modality is also used as input to activate the encoder normalizations corresponding to the data type and generate common latent spaces.}
    \label{fig:intro}
\end{figure}

Medical image segmentation is one of the most challenging and studied research fields of deep learning and artificial intelligence in recent years \cite{siddique_u-net_2021, suganyadevi_review_2022}. The precision of segmentation achieved competitive results thanks to recent efforts in developing convolution \cite{malhotra_deep_2022, ronneberger_u-net_2015} and transformer-based neural networks \cite{hatamizadeh_swin_2022, hatamizadeh_unetr_2021}. These algorithms are generally trained on a single medical imaging modality, such as T1- or T2-Weighted Magnetic Resonance Images (MRIs) or Computed Tomography (CT) scans. Therefore, they often suffer from data variability when tested on images different from those seen during training. Data variability is a common issue in the medical field and it is due to several uncontrollable factors that depend on the specific clinic or patient needs \cite{suetens_fundamentals_2017}, e.g., different imaging methods, different scanners, different acquisition settings or patient limitations. In this scenario, the most straightforward approach would be to train a separate model for each domain, requiring a massive amount of different annotated training data and not exploiting inter-domain information.

Recently, to tackle this issue, some multi-modal \cite{wang_multimodal_2022} and cross-modality \cite{valindria_multi-modal_2018, li_towards_2020, zhang_translating_2019, zhou_generalizable_2022} medical image segmentation techniques have been proposed. The first group of methods aims to produce better segmentation by simultaneously exploiting information from multiple sources. In this case, it is good practice to stack several images of different types to generate a combined input for the segmentation model \cite{zhou_review_2019}. Furthermore, multi-modality can also be leveraged to produce synthetic images of missing domains at inference time \cite{dalmaz_resvit_2022}.  Here, different modalities are rather seen as separate inputs for the model that generates a common representation and produces synthetic images. However, in both scenarios, registered medical images \cite{hill_medical_2000} are needed. Although registration is relatively simple, the acquisition of several images of the same patient is often limited by local resources and time constraints. 

The second group of methods, i.e. cross-modality methods, focus on producing segmentation independently of the data type provided in input. More specifically, an \textit{assistant modality}, of which several annotated data are available, is often used to improve the segmentation performance on another modality with fewer annotated data, i.e. \textit{target modality}. Fine-tuning \cite{tajbakhsh_convolutional_2016} or transfer learning \cite{raghu_transfusion_2019} are straightforward ways of adapting deep learning models to target modalities after they have been trained on some specific assistant modality. However, these approaches do not fully exploit shared cross-modality information. One intuitive way to overcome this problem is joint training \cite{valindria_multi-modal_2018}, where a model is trained simultaneously with multiple modalities. Nevertheless, it is relatively hard to directly learn common features when the domain shift is significant, e.g MRIs and CTs. Therefore, to directly learn from multi-modality data with large appearance discrepancy, the use of several feature extractors, i.e., one per modality, has been first proposed in an X-shaped architecture \cite{valindria_multi-modal_2018}, implying a significant overhead on the segmentation model and requiring adjustment to generalize to other clinical tasks. To address this issue, several techniques based on synthetic image generation prior to segmentation have been introduced \cite{chen_crdoco_2020, jiang_tumor-aware_2018, zhou_generalizable_2022}. In many cases, the assistant modality is used to synthesize the target, or vice versa, and train the model on a mix of real and synthetic data \cite{li_towards_2020}. The additional pre-processing step required by these methods to make the clinical images segmentation-ready adds computational complexity, which is often critical in real-time applications such as computer-assisted surgery \cite{tanzi_real-time_2021}.

In this work, we propose a general framework, applicable to any encoder-decoder literature architecture, aiming to produce high-quality cross-modality segmentation without introducing overhead on the segmentation model nor needing registered clinical images for training. With our framework, medical images from different domains can be fed individually directly into a single modality-conditioned model, which generates the desired segmentation by self-adapting its encoder normalization layers. The adaptation is based on Conditional Instance Normalization (CIN), which has been originally proposed for the transfer of arbitrary artistic styles by Dumoulin \textit{et al.} \cite{dumoulin_learned_2017}. Thanks to that, and similarly to joint training, our models can be trained in an end-to-end way by randomly mixing different data of several modalities, in what we call \textit{interleaved mixed training} fashion, without prior synthetic image style transfer. Furthermore, based on the proposed framework, we formally define the Conditional Vision Transformer (C-ViT) encoder architecture to build modality-agnostic ViT-based models \cite{dosovitskiy_image_2021} for image segmentation or classification, as shown in Figure \ref{fig:intro}. We extensively evaluate our method on the Multi-Modality Whole Heart Segmentation (MM-WHS) 2017 Challenge \cite{zhuang_evaluation_2019}. We show that C-ViT improves the baseline model and other multi-modal techniques and that our framework outperforms other cross-modality segmentation methods, when applied to the same baseline model. Our contributions are summarized as follows:

\begin{itemize}[noitemsep,topsep=0pt]
    \item We present a novel framework to enhance the quality of cross-modality segmentation, while removing expensive input pre-processing steps and the need of registered data for training.
    \item Following the framework, we introduce the Conditional Vision Transformer encoder, which adapts its normalization layers based on the input modality and can be used for several tasks.
    \item  We evaluate our framework on the public whole heart and substructures segmentation challenge \cite{zhuang_evaluation_2019}, showing that it outperforms other cross-modality learning techniques both on UNet and ViT baselines.
\end{itemize}

\section{Related Works}
\textbf{Multi-Modality Learning in Medical Imaging.} In recent years, several deep learning architectures have been proposed for image segmentation, achieving extremely good performance \cite{chen_transunet_2021, hatamizadeh_swin_2022, hatamizadeh_unetr_2021, milletari_v-net_2016, sengara_unet_2022}. Among them, UNet \cite{ronneberger_u-net_2015} is surely the most popular and is used as a baseline to create better performing models \cite{siddique_u-net_2021}. More recently, some variation based on vision transformers (ViT) \cite{dosovitskiy_image_2021} and swin-transformers \cite{liu_swin_2021}, such as TransUNET \cite{chen_transunet_2021}, UNETR \cite{hatamizadeh_unetr_2021} and Swin-UNETR \cite{hatamizadeh_swin_2022}, have been shown to outperform previous versions of UNet. In the clinical imaging field, multi-modality learning is a similar, yet challenging, task. Multiple medical image domains have been exploited for synthetic image generation based on Generative Adversarial Networks (GAN) \cite{dalmaz_resvit_2022, dar_image_2019, goodfellow_generative_2014, sun_synthesis_2022, zhang_multi-contrast_2022} or multi-modal image segmentation \cite{fidon_scalable_2017, guo_medical_2018, wang_multimodal_2022, zhou_review_2019}. Interestingly, the two tasks have also been merged by Zhang \textit{et al.} \cite{zhang_cross-task_2022} in a cross-task feedback fusion GAN that first generates synthetic CT and then performs multi-modal segmentation. The latter uses cross-domain information to achieve better performances, but it is restricted to when registered images are available. To relax the need of aligned medical images and exploit inter-domain features during training, cross-modality segmentation approaches have recently gained attraction. Firstly, Zheng \textit{et al.} \cite{zheng_cross-modality_2015} investigated the effectiveness of shape priors learned from an assistant modality to improve the segmentation on a target modality by marginal space learning. Valindria \textit{et al.} \cite{valindria_multi-modal_2018} developed dual-stream encoder-decoder models having different branches for each modality and implementing weights sharing techniques to extract cross-modality features. Similarly to our work, some researcher tried to play with the normalization layers of the models to improve their generalization capacities \cite{fan_adversarially_2021, pan_two_2020, segu_batch_2021, seo_learning_2020, zhou_generalizable_2022}. Among others, Pan \textit{et al.} \cite{pan_two_2020} introduced the IBN-Net which simultaneously exploiting Instance and Batch Normalization to capture both appearance changes and content information. Segu \textit{et al.} \cite{segu_batch_2021} proposed to collect prior domain-dependent statistics by training ad-hoc Batch Normalization (BN) to map the modalities on a shared latent space. 

With the aforementioned advances in synthetic images generation, many works tried to tackle the challenge by assisting segmentation models with prior image translation \cite{chen_crdoco_2020, hoffman_cycada_2017, jiang_tumor-aware_2018, li_towards_2020, zhang_translating_2019, zhou_generalizable_2022}. Among them, Zhang \textit{et al.} \cite{zhang_translating_2019} tried to improve the segmentation for modalities with limited training samples by easing the appearance gap with other modalities using a GAN. Similarly, Li \textit{et al.} \cite{li_towards_2020} proposed an Image Alignment Module to reduce the appearance gap between assistant and target modality and introduced a Mutual Knowledge Distillation scheme \cite{hinton_distilling_2015} to exploit modality-shared knowledge. More recently, Zhou \textit{et al.} \cite{zhou_generalizable_2022} proposed to simulate the possible appearance changes of a target domains by non-linear transformation to augment source-similar and source-dissimilar images. 

Different from these methods, which mainly rely on image translation or prior trainings, our framework tries to shrink the model complexity and, at the same time, improve the segmentation accuracy extracting all the available cross-modality features by conditioning the segmentation models and training them jointly with assistant and target modality. Moreover, many of the previous techniques do not allow for arbitrary switch of the input modality, but they just exploit assistant modality information to improve the target segmentation.

\begin{figure}[t]
    \centering
    \includegraphics[width=0.48\textwidth]{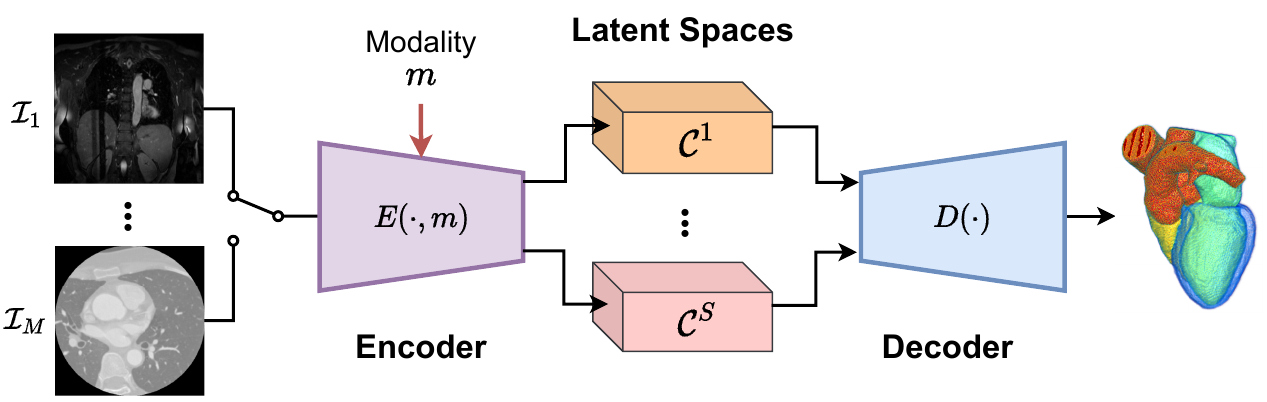}
    \caption{Overview of the proposed conditional framework: the conditional encoder, $E(\cdot, m)$, generates $S$ common latent spaces, $\{ \mathcal{C}^1, \dots, \mathcal{C}^S\}$, one for each encoder-decoder connection,  adapting itself to the input modality $m$ and the decoder, $D(\cdot)$, is unique and does not need adaptation.}
    \label{fig:conditional_model}
\end{figure}

\textbf{Normalization Layers.} Our work is also related with normalization layers used in neural networks, which may differ based on the spatial dimension to which they are applied. The four main classes of normalization layers are batch, layer, group, and instance normalization. Batch Normalization was the first technique introduced to deal with the variability of layers input during training by normalizing the features across mini-batches \cite{ioffe_batch_2015}. Layer Normalization (LN) \cite{ba_layer_2016}, used mainly in Natural Language Processing (NLP) and Recurrent Neural Networks (RNNs), was introduced to relax the dependency on the batch, normalizing only across features. Similarly, Group Normalization \cite{wu_group_2018} computes the mean and standard deviations in groups of channels rather than using the entire feature space. Finally, Instance Normalization (IN) has been introduced, in the imaging context, to prevent instance-specific mean and covariance shift, simplifying the learning process \cite{ulyanov_instance_2017}. Huang \textit{et al.} \cite{huang_arbitrary_2017} proposed a further variation of IN, called Adaptive IN to perform arbitrary image style transfer, that is, synthetic image generation, by adjusting the mean and variance of the content input to match those of the desired style input. The latter does not have learnable parameters, but it computes them directly from the input style. In our framework, we propose to use a similar variation of IN recently proposed by Dumoulin \textit{et al.} \cite{dumoulin_learned_2017}, called Conditional Instance Normalization (CIN), which learns a different set of parameters $\gamma_m$ and $\beta_m$ for each style $m \in [1,M]$. Given a three-dimensional volume with $C$ channels, $\pmb{z} \in \mathbb{R}^{ C\times H\times W\times D}$, of modality $m$, the CIN is calculated as 
\begin{equation}
\label{eq:CIN}
    \text{CIN}(\pmb{z}, m) = \gamma_m \left(\frac{\pmb{z} - \mu(\pmb{z})}{\sigma(\pmb{z})} \right) + \beta_m
\end{equation}
where $\mu(\pmb{z})$ and $\sigma(\pmb{z})$ are computed across spatial dimensions independently for each channel and each sample \cite{huang_arbitrary_2017}. It allows the sharing of all convolutional weights of a style transfer network across many image styles by only changing the learnable normalization parameters. Inspired by that, in this work, CIN is exploited to generate a shared latent space for different input modalities, which is used to obtain the segmentation output independently of the medical input image domain.

\section{Method}
The overview of the proposed cross-modality segmentation framework is shown in Figure \ref{fig:conditional_model}. The segmentation model, which can be chosen among those available in the literature, is conditioned by the input data modality to adapt its normalization layers to the fed clinical image. Therefore, the aim is to produce the desired segmentation independently of the input data domain. 

In the following, the general framework for creating conditional models is introduced and it is applied to the transformer encoder, to obtain the C-ViT encoder. Finally, more details about the framework training procedure, including the loss function, are reported. 

\textbf{Conditional Segmentation Models.} Let any state-of-the-art encoder-decoder segmentation architecture be considered as baseline for the proposed framework. In a straightforward case, for a given data modality $m \in [1,M]$, let $E_m(\cdot): \mathcal{I}_m \rightarrow \mathcal{H}_m$ and $D_m(\cdot): \mathcal{H}_m \rightarrow \mathcal{O}$ be the two functions that describe the encoder and decoder modules of the chosen architecture, respectively. The role of $E_m(\cdot)$ is to transform the input, $\pmb{x}_m \in \mathcal{I}_m$, into a latent feature space, $\mathcal{H}_m$, which can better represent the information carried by raw data. On the other hand, the purpose of $D_m(\cdot)$ is to transform this representation into the final segmentation $\pmb{y} \in \mathcal{O}$. In general, the space $\mathcal{H}_m$ is learned ad-hoc by the model, based on the type of data used during training. 

To make the model independent of the input modality, we propose to create a general conditional encoding module, $E(\cdot, m): \mathcal{I}_m \rightarrow \mathcal{C}$, which transforms any input $\pmb{x}_m \in \mathcal{I}_m$, $1\le m \le M$, into a common shared latent feature space $\mathcal{C}$. In this way, a single decoder $D(\cdot): \mathcal{C} \rightarrow \mathcal{O}$ is needed to generate the segmentation of all the modalities. The relation between any input $\pmb{x}_m$ and the output $\pmb{y}$ in a conditional model can then be expressed as 
\begin{equation}
\label{eq:conditional_models}
    \pmb{y} = D(E(\pmb{x}_m, m)) \;,\;  {1 \le m \le M}
\end{equation}

This concept is easily extended to architectures with skip connections, e.g. UNet \cite{ronneberger_u-net_2015}, in which the encoder may be seen as a sequence of $S$ sub-modules, each of them generating a common latent space $\mathcal{C}^s$, $0\le s\le S$, where $S$ is also the number of connections between encoder and decoder. Therefore, the decoder will simply gather the data from all the shared latent spaces to generate a modality-agnostic segmentation output.

In practice, we propose to generate the conditional encoder $E(\cdot, m)$, starting from any simple encoder $E_m(\cdot)$, by adding the data modality as input (often available from meta-data, without manual intervention) and replacing all its normalization layers with CIN. In such way, the only overhead on the segmentation model is given by the extra normalization learnable parameters $\gamma_m$ and $\beta_m$, $1\le m \le M$, of Equation \ref{eq:CIN}, which are negligible with respect to the rest of the architecture. Furthermore, we avoid any pre-processing phase, such as synthetic image generation through GANs, which can heavily impact on the network performances, and we let the model directly learn a common shared latent space for all the input modalities, trying to fully exploit the inter-modality information.

\begin{figure}[t]
    \centering
    \includegraphics[width=0.47\textwidth]{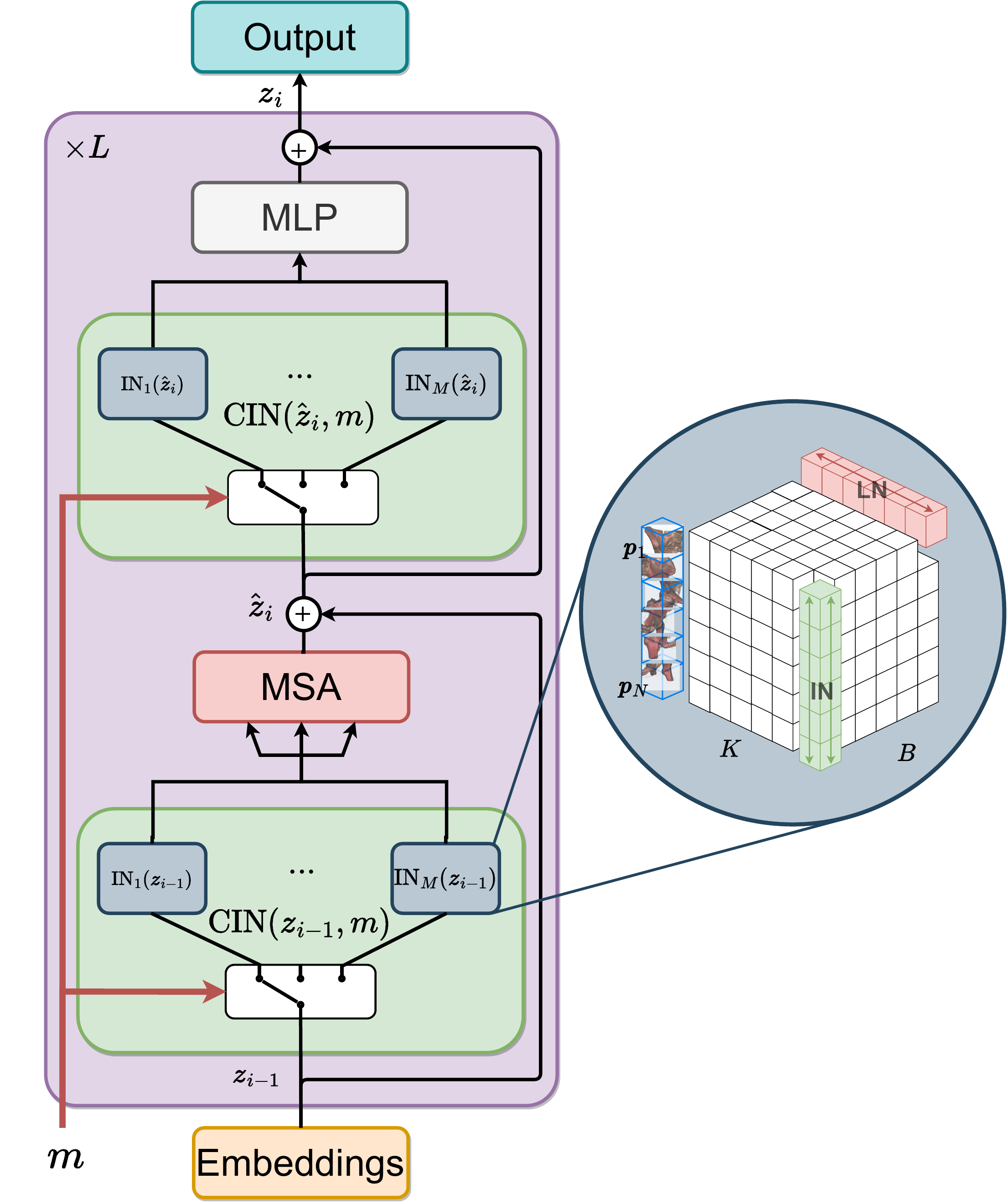}
    \caption{Conditional Vision Transformer Encoder. The multi-head self-attention (MSA) mechanism is followed by a MLP and each of them is preceded by CIN. The latter consists of a switch mechanism to activate the IN corresponding the to modality flag in input. A focus on the different normalizing direction of IN and LN is also shown in the zoom.}
    \label{fig:transformer_block}
\end{figure}

\begin{table*}[t]
\caption{Quantitative comparison with other methods for cross-modality medical image segmentation on the target modality (CT). All the techniques have the same UNet basline \cite{valindria_multi-modal_2018}, are trained using 20 MRIs and 10 CTs and are evaluated on the test set of MM-WHS dataset. The mean DICE score is reported, as well as the ones for all the heart substructures.}
    \centering
\begin{tabularx}{\textwidth}{l|Y|Y|Y|Y|Y|Y|Y|Y}
\toprule
 \multirow{2}{*}{Model} & \multirow{2}{*}{Avg Dice} & \multicolumn{7}{c}{Dice of substructures of heart}\\
 \cline{3-9}
 & & MYO & LA & LV & RA & RV & AA & PA\\
 \midrule
 Baseline & 87.06 & 87.02 & 89.22 & 90.86 & 83.86 & 84.60 & 92.52 & 81.34 \\
 Fine-tune & 87.69 & 87.16 & 90.40 & 90.79 & 84.43 & 85.26 & 92.74 & 83.05 \\
 Joint-training & 87.43 & 86.65 & 90.76 & 91.23 & 82.78 & 84.92 & 93.02 & 82.66 \\
 X-shape \cite{valindria_multi-modal_2018} & 87.67 & 87.19 & 89.79 & 90.94 & 85.51 & 84.44 & 93.43 & 82.40 \\
 Zhang \textit{et al.}\cite{zhang_translating_2019} & 88.50 & 87.81 & 91.12 & 91.34 & 85.14 & 86.31 & 94.30 & 83.42 \\
 Li \textit{et al.} \cite{li_towards_2020} & 90.12 & 89.34 & 91.90 & 92.67 & 87.47 & 88.14 & \textbf{95.95} & \textbf{85.38} \\
 \textbf{Ours} & \textbf{90.77} & \textbf{90.06} & \textbf{92.68} & \textbf{93.77} & \textbf{88.22} & \textbf{90.85} & 94.70 & 84.52\\
\bottomrule
\end{tabularx}
    \label{tab:Dice_comparison}
\end{table*}

\textbf{Conditional Vision Transformer.}
The proposed C-ViT encoder is shown in Figure \ref{fig:transformer_block}. As originally proposed by Vaswani \textit{et al.} \cite{vaswani_attention_2017}, it has two sub-layers, the ﬁrst is a multi-head self-attention mechanism (MSA), and the second is a simple, position-wise fully connected Multi-Layer Perceptron (MLP). Both sub-layers are preceded by a normalization \cite{dosovitskiy_image_2021, wang_learning_2019} and a residual connection is employed around each of them. Here, we replace the Layer Normalization (LN) \cite{ba_layer_2016}, which has historically been used in transformers, with CIN. In other words, a switch mechanism that selects the IN corresponding to the input modality is introduced before each sub-layer. 

In a general scenario with $M$ different input modalities, let $L$, $N$ and $K$ be the number of stacked transformer blocks that make up a transformer encoder, the number of input patches and the dimensions of the embedding space, respectively. For a given input embedding of the $i$-th transformer block, $\pmb{z}_{i-1} \in \mathbb{R}^{N \times K}$, we have
\begin{equation}
\label{eq:MSA}
    \hat{\pmb{z}}_{i} = \text{MSA}(\text{CIN}(\pmb{z}_{i-1}, m)) + \pmb{z}_{i-1} \;,\;  {1 \le i \le L}
\end{equation}
\begin{equation}
    \pmb{z}_{i} = \text{MLP}(\text{CIN}(\hat{\pmb{z}}_{i}, m)) + \hat{\pmb{z}}_{i} \;,\;  {1 \le i \le L}
\end{equation}
where $1 \le m \le M$ is the modality corresponding to the input and $\text{CIN}(\cdot, \cdot)$ is defined in Equation \ref{eq:CIN}. Note that $\pmb{z}_{0}$ corresponds to the linear projection of the input patches, i.e. non-overlapping sub-patches of dimension $P\times P \times P$ of the input volume, $\{\pmb{p}_j \in \mathbb{R}^{P^3} : 1 \le j \le N\}$ , into a $K$-dimensional embedding space using a projection matrix $\pmb{E} \in \mathbb{R}^{P^3 \times K}$ \cite{hatamizadeh_unetr_2021}. Additionally, learnable positional embeddings, $\pmb{E}_{\text{pos}} \in \mathbb{R}^{N \times K}$, are included to code spatial information, finally obtaining
\begin{equation}
    \pmb{z}_{0} = [\pmb{p}_1 \pmb{E}; \dots ; \pmb{p}_J \pmb{E}] + \pmb{E}_{\text{pos}}.
\end{equation}

The rationale for substituting LN, which is inherited from the applications of the first transformers in NLP \cite{wolf_huggingfaces_2020}, for CIN, or more generally IN, is also motivated on the zoom of Figure \ref{fig:transformer_block}. LN computes the average and variance of the single input patches along the $K$ dimensions of the embedding space. Therefore, when it comes to images, the latter produces an independent normalization for each image patch without considering the entire input data. IN addresses this issue by computing the metrics for normalization across all the input patches but separately for each embedding dimension. Intuitively, the model is able to generate more meaningful embeddings when the input volume is considered as a whole for normalization.

The technique presented here can also be generalized to the \textit{Swin-transformer} model by replacing the MSA of Equation \ref{eq:MSA} with W-MSA or SW-MSA, denoting window-based multi-head self-attention using regular and shifted window partitioning configurations, respectively \cite{liu_swin_2021}. Finally, in this case, to have a fully conditional encoder, also the patch merging block, used to generate hierarchical representations of the input patches, must be adapted in the normalizations. 

\textbf{Loss function and Training.} To train cross-modality segmentation models using our framework, we used a combination of Dice Loss \cite{jadon_survey_2020}, $\mathcal{L}_\text{Dice}$, and Focal Loss \cite{lin_focal_2018}, $\mathcal{L}_\text{Focal}$, as
\begin{equation}
\label{eq:loss}
    \mathcal{L} = \lambda_D \mathcal{L}_\text{Dice} + \lambda_F \mathcal{L}_\text{Focal}
\end{equation}
where $\lambda_D$ and $\lambda_F$ are scalar factors greater than $0$ to adjust the contribution of each individual loss to the total. We preferred this version of the loss, rather than the standard combination of Dice loss and Cross-Entropy (CE), to tackle the unbalancing of clinical image segmentation, which is often significant between background and foreground. The focal loss, instead, reshapes the standard CE loss so that it down-weights the loss assigned to well-classified examples. In this way, more weight will be given in the loss to hard classified samples, i.e., labels difficult to segment, improving the learning process. 

Differently from joint training \cite{valindria_multi-modal_2018}, in which training data is divided into batches of different modalities, conditional models are trained using an interleaved mixed data training fashion. This means that different modalities are randomly fed into the model during training, even in the same batch. Therefore, the overall loss will provide an overview of the goodness of the segmentation independently from the input modality, pushing the network to be fair with all different types of data.

\begin{table*}[t]
\caption{Performance comparison on the target modality (CT) of our framework applied to the Swin-UNETR model \cite{hatamizadeh_swin_2022}. All the models are trained using 20 MRI and 10 CT volumes and are evaluated on the test set of MM-WHS dataset. The mean DICE score is reported, as well as the ones for all the heart substructures and Whole Heart (WH) segmentation.}
    \centering
\begin{tabularx}{\textwidth}{l|Y|Y|Y|Y|Y|Y|Y|Y|Y}
\toprule
 \multirow{2}{*}{Model} & \multirow{2}{*}{WH} & \multirow{2}{*}{Avg Dice} & \multicolumn{7}{c}{Dice of substructures of heart} \\
 \cline{4-10}
 & & & MYO & LA & LV & RA & RV & AA & PA\\
 \midrule
 Baseline & 83.11 & 85.32 & 84.79 & 88.12 & 89.80 & 81.04 & 84.08 & 77.58 & 76.36\\
 Fine-tune & 86.33 & 86.09 & 82.57 & 90.03 & 87.92 & 83.29 & 85.51 & 90.28 & 83.06\\
 Joint-training & 89.03 & 87.99 & 86.25 & \textbf{92.07} & 91.97 & 85.35 & 87.98 & 89.55 & 82.76\\
 \textbf{C-ViT} [ours] & \textbf{89.98} & \textbf{89.33} & \textbf{88.13} & 91.64 & \textbf{92.39} & \textbf{86.36} & \textbf{89.33} & \textbf{93.39} & \textbf{84.05}\\
\bottomrule
\end{tabularx}
    \label{tab:Dice_swin}
\end{table*}

\begin{table}[t]
\caption{Performance comaprison on MRIs, used as assistant modality, of our framework applied to the Swin-UNETR model \cite{hatamizadeh_swin_2022}. The baseline is the model used to do fine-tuning on the target modality.}
    \centering
\begin{tabularx}{0.4\textwidth}{l|Y|Y}
\toprule
 Model & WH & Mean Dice\\
 \midrule
 Baseline&  83.32 & 81.87\\
 Joint-training & 86.65 & 84.82\\
 C-ViT [ours] & 85.57 & 84.30\\
\bottomrule
\end{tabularx}
    \label{tab:Dice_mri}
\end{table}

\section{Experiments}
In this section we describe the experiments carried out to evaluate the proposed cross-modality framework with interleaved mixed data training. The experiment settings are described, including the dataset, the implementation details, and the algorithms to tune the models hyper-parameters. Finally, the results are shown and discussed.

\subsection{Experiment settings}
\textbf{Dataset.} We evaluate the proposed method on the Multi-modality Whole Heart Segmentation Challenge 2017 (MM-WHS 2017) dataset \cite{zhuang_evaluation_2019}, which contains non-registered 20 MRI and 20 CT volumes for training and the ground-truth (GT) annotations of 7 cardiac substructures including the Left Ventricle blood cavity (LV), the Right Ventricle blood cavity (RV), the Left Atrium blood cavity (LA), the Right Atrium blood cavity (RA), the MYOcardium of the left ventricle (MYO), the Ascending Aeorta (AA), and the Pulmonary Artery (PA). Following previous cross-modality segmentation works on this dataset \cite{li_towards_2020}, we used MRI as assistant modality and CT as target modality, since MRI has better contrast for soft tissue and ideally provide better information for heart substructures segmentation. We evenly and randomly split the CT data in order to perform a two-fold cross-validation and train the conditional models with 20 MRIs and 10 CTs at a time, to simulate data shortage for the target modality. Furthermore, MRI volumes are pre-processed using the N4 bias field correction algorithm to correct low frequency intensity non-uniformity present in MRI image data known as gain field \cite{tustison_n4itk_2010}. Finally, all the volumes are re-sampled to an isotropic space and their values are normalized between 0 and 1.

\textbf{Implementation Details.} To provide a fair comparison, we first applied our framework to the same baseline used by other cross-modality previous works \cite{li_towards_2020, valindria_multi-modal_2018}. Namely, we implemented the network architecture with encoder-decoder structure proposed by Valindria \textit{et al.} \cite{valindria_multi-modal_2018}, consisting of a 3D UNet \cite{ronneberger_u-net_2015} with residual blocks with pre-activation \cite{he_identity_2016}. After that, we evaluated the efficiency of the proposed C-ViT encoder applied to the Swin-UNETR \cite{hatamizadeh_swin_2022} segmentation model, originally consisting of a Swin-transformer encoder \cite{liu_swin_2021} and UNet style decoder, as shown in Figure \ref{fig:intro}.  For all the models, we define an input size of $96 \times 96 \times 96$ pixels and sub-patches, in the case of transformer-based architectures, of $16 \times 16 \times 16$ pixels. Therefore, at each epoch, a fixed number of random crops of input size is extracted, for each volume. Note that, in this way, the effective batch size is given by $N_s \times N_c$, where $N_s$ and $N_c$ are the number of samples and crops, respectively. For the transformer-based models, we follow the original implementation of Hatamizadeh \textit{et al.} \cite{hatamizadeh_swin_2022}, with $L=8$ blocks and hidden size $K=768$ in the encoder. Furthermore, data augmentation is applied to artificially increase training data and deal with inter-modality variations by randomly shifting and scaling the intensity of the input volumes, along with random flips in all three dimensions. Finally, to validate and test the models, sliding window inference is used with an overlap of $0.5$.

The code is implemented in PyTorch, with the help of the MONAI library \cite{cardoso_monai_2022}, and trained on a cluster equipped with multiple NVIDIA V100 GPUs. Optimization is performed using Adam with decoupled weight decay \cite{loshchilov_decoupled_2019} and a cosine-decay Learning Rate (LR) scheduler with warm-up, for a maximum of 2500 epochs. All the models are tuned to find the best set of hyper-parameters as described below. 

\begin{figure*}[t]
    \centering
    \includegraphics[width=.80\textwidth]{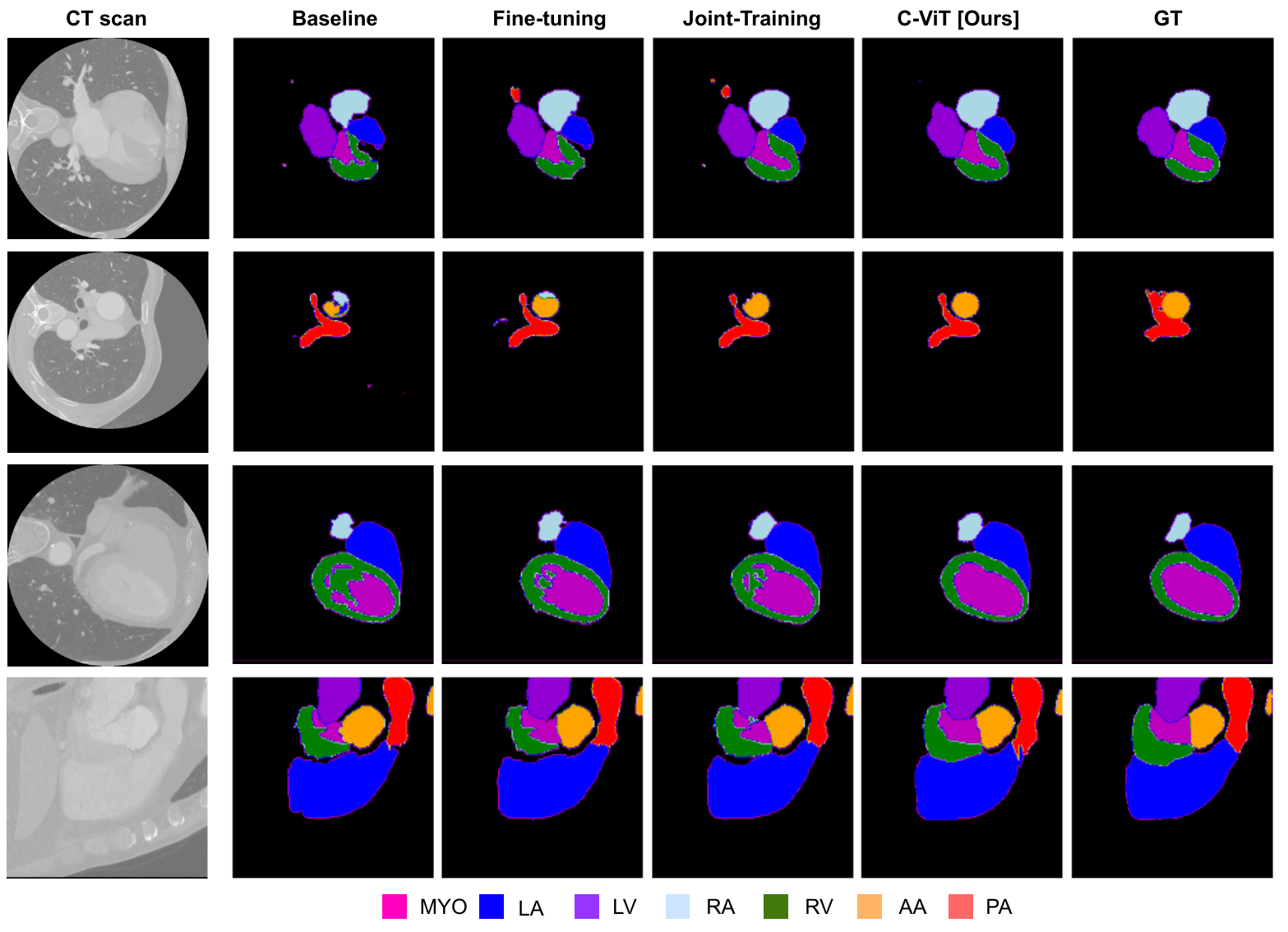}
   \caption{Qualitative comparison of the segmentation result on the CT target domain of our framework applied to the Swin-UNETR model \cite{hatamizadeh_swin_2022} on the MM-WHS dataset. Our framework is compared with the baseline model, trained on the target modality, fine-tuning and joint-training with interleaved mixed fashion.}
\label{fig:results}
\end{figure*}

\textbf{Hyper-parameters Tuning.} To take full advantage of the segmentation models, we tuned them using the Optuna framework \cite{akiba_optuna_2019}, following the guidelines from \cite{godbole_deep_2023}. Therefore, the batch size is not tuned but is kept unchanged among trials as the maximum that can fit our memory availability, that is, 16 three-dimensional crops ($N_s = 4$ and $N_c = 4$). The optimal LR is found for each model using the cyclical learning rate technique \cite{smith_cyclical_2017} and a logarithmic search space centered on that value is defined for tuning. The number of LR Warm-Up epochs is chosen among the values of 2, 4 and 6\% of the maximum training epochs \cite{izsak_how_2021} and the optimizer weight decay is also tuned. Finally, the loss function of Equation \ref{eq:loss} is used with $\lambda_D = \lambda_F = 1$. 

For each trial, the hyper-parameters are sampled form the search space using the Tree-structured Parzen Estimator (TPE) algorithm \cite{bergstra_algorithms_2011, ozaki_multiobjective_2022}. This technique aims to find the best set of parameters by fitting two Gaussian Mixture Models (GMMs) to the good and bad trials and minimizing their ratio. Furthermore, Asynchronous Successive Halving (ASHA) \cite{li_system_2020} is used to early-stop bad trials and save computational time and resources.

\subsection{Experiment results} 
We report here the results of our experiments using the proposed framework on the MM-WHS dataset \cite{zhuang_evaluation_2019}, quantitative using the Dice similarity coefficient and qualitative by comparing segmentation images. 
In the following, we first compare our proposal with other techniques and then we evaluate the accuracy of the proposed C-ViT. 

\textbf{Comparison with Other Methods.} The comparative quantitative results for heart substructures segmentation are shown in Table \ref{tab:Dice_comparison}. In fine-tuning the model is first trained using the assistant modality only and then, the acquired knowledge is transferred to the target modality with a subsequent training. In join-training the baseline model is trained with both modalities simultaneously by alternating them on different batches. We also compare the accuracy with the X-shape architecture proposed by Valindria \textit{et al.} \cite{valindria_multi-modal_2018} and the online training with synthetic image generation approach of Zhang \textit{et al.} \cite{zhang_translating_2019}. Finally, a knowledge distillation cross-modality segmentation technique is also considered for comparison \cite{li_towards_2020}. As highlighted by the latter, fine-tuning, joint-training and the X-shaped architecture bring marginal improvements on the segmentation accuracy since they do not fully exploit cross-modality information. Online GAN-based synthetization with and without mutual knowledge distillation start to significantly improve the mean accuracy of target modality segmentation, up to $3.06\%$. Nevertheless, in both cases, significant overhead is introduced on the model, limiting its possible real-time applications. Our method achieves significant improvements in the segmentation accuracy, while reducing the complexity of both training and inference. We observe an enhancement of $0.65\%$ in the mean Dice and up to $2.71\%$ on the single heart substructures, namely in the RV. However, a small decrease in performances is registered for AA and PA, probably due to the fact that the manual segmentation of these structures generally covers beyond the real size of the vessels while the test tool cuts them to a limited length \cite{zhuang_evaluation_2019}.
\begin{figure}[t]
    \centering
    \includegraphics[width=.46\textwidth]{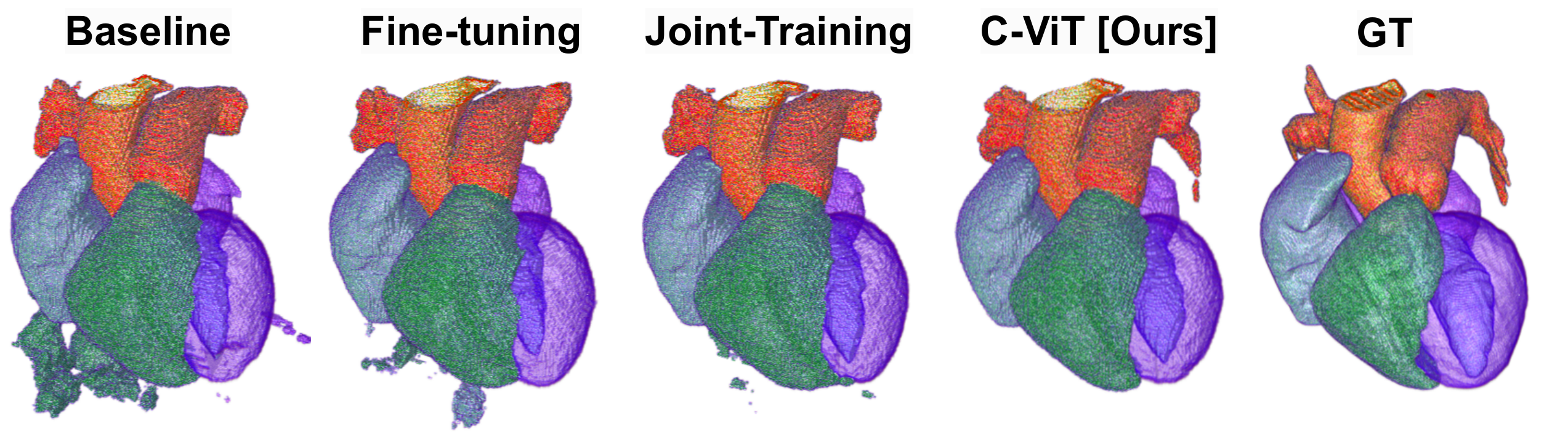}
   \caption{Comparison of the 3D segmentation of heart substructures from CT using different cross-modality adaptations.}
\label{fig:results_3d}
\end{figure}
\begin{figure}[t]
    \centering
    \includegraphics[width=.48\textwidth]{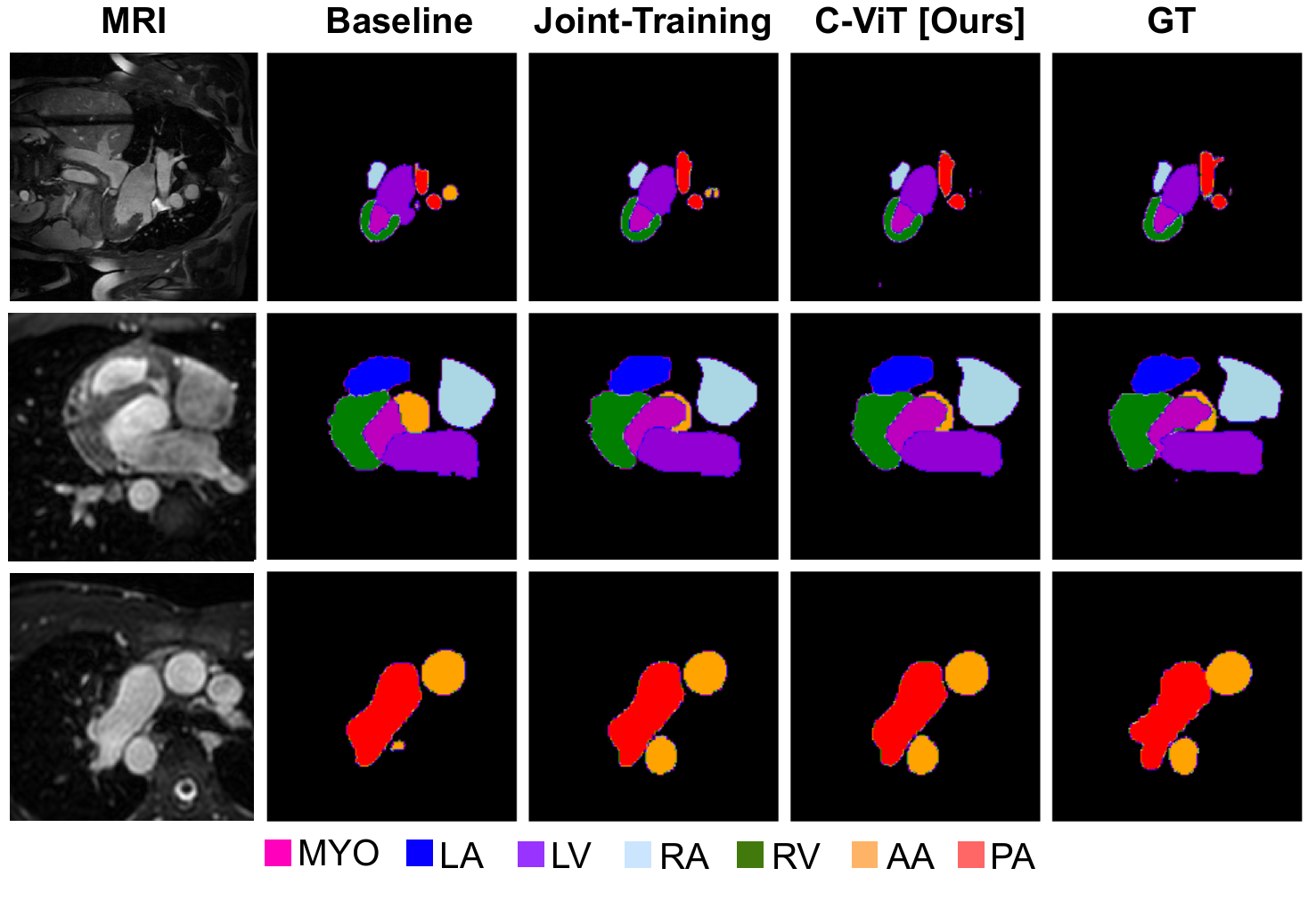}
   \caption{Qualitative comparison of the segmentation result, on the MRI assistant domain, of our framework applied to the Swin-UNETR model \cite{hatamizadeh_swin_2022} on the MM-WHS dataset. }
\label{fig:results_mri}
\end{figure}

\textbf{Evaluation of C-ViT and Ablation Study.} We compare the performance of the C-ViT-based segmentation model with its baseline \cite{hatamizadeh_swin_2022} and the result obtained by fine-tuning and joint training the same model (with interleaved mixed fashion). Table \ref{tab:Dice_swin} summarizes the quantitative results in terms of Dice accuracy of Whole Heart Segmentation (WHS) and single substructures segmentation. Furthermore, Figure \ref{fig:results} shows a qualitative comparison on single slices of the segmentation obtained on CT by using different methods, and Figure \ref{fig:results_3d} directly compares the 3D segmentation of a whole sample. Our framework enhances of $4\%$ the mean Dice of heart substructures, with significant improvements in each of them, and $6.87\%$ the accuracy of WHS, with respect to the baseline model. Qualitative, we can see important improvements when using C-ViT with respect to other methods, that may also result in refinement of the GT (e.g. the segmentation of PA on the second row of Figure \ref{fig:results}). In the 3D comparison, we observe the improvement in the segmentation produced by C-ViT which does not contain false positives outside the region of the GT. Although in our cross-validation the ViT-based conditional model performs better than the UNet one ($91.18\%\pm0.68$ vs $90.76\%\pm0.55$ of mean Dice), in the test set, the latter seems to be slightly better likely due to the fact that transformers usually need more data for finer generalization \cite{park_how_2022}.

Differently from some other related works \cite{jiang_tumor-aware_2018, li_towards_2020} our framework do not just exploit assistant modality to improve the segmentation of the target one, but it keeps the ability of producing high-quality segmentation for the assistant modality by switching the input domain. To confirm that, In Table \ref{tab:Dice_mri}, WHS and Mean Dice of MRI segmentation is reported, where the baseline is considered to be the model trained only with assistant modality and used to fine-tune the target modality. Also the assistant modality segmentation accuracy benefits of our framework, as well as of joint-training, improving with respect to the baseline model. Moreover, Figure \ref{fig:results_mri} also shows some qualitative improvements of the MRI segmentation, especially from baseline to C-ViT. As previously mentioned, transformers are "\textit{data-hungry}" \cite{wang_towards_2022}, in the sense that they suffer from significant performance drops on small-size datasets. We have shown that, with our conditional framework applied to ViT, it is possible to relax the need of an huge dataset for training by gathering several small datasets of different modalities. 

\section{Conclusions}
In this work, we proposed a simple framework that aims to segment different types of medical images, while reducing the model overhead and the need of registered data for training, using a single cross-modality conditional model trained with interleaved mixed data. We introduced a general definition of a conditional model, based on CIN, that can be applied to all state-of-the-art medical image segmentation architectures. Furthermore, we developed a new C-ViT encoder that can be used to create conditional ViT-based models. We validated our framework using a public multi-modal dataset for heart substructures segmentation, which collects both MRI and CT data. Our framework achieves new state-of-the-art performance for the cross-modality medical image segmentation with assistant and target modality. We have shown that not only the first one helps the learning on the latter when few annotated data are available, but also the segmentation of the assistant modality benefits of our framework. We believe that our work can be an important contribution for developing robust cross-modality medical image segmentation methods, to tackle cases in which the acquisition of a specific kind of image is limited by environmental or patient-specific restrictions. Furthermore, in the future, the proposed framework may be extended in an Unsupervised Domain Adaptation (UDA) fashion, to make possible the training on a single labeled modality with adaptation to unlabeled domains.

\section{Acknowledgements}
This project has received funding from the European Union’s Horizon 2020 research and innovation programme under the Marie Skłodowska-Curie grant agreement No 945304-Cofund AI4theSciences hosted by PSL University.

This work was granted access to the HPC/AI resources of IDRIS under the allocation 2022-AD011013902 made by GENCI.

{\small
\bibliographystyle{ieee_fullname}
\bibliography{references}
}

\end{document}